   \documentclass[prc, 12pt, nofootinbib]{revtex4}
\usepackage{epsfig,amssymb,bm,graphics,color}
\usepackage[caption=false]{subfig}
\def\ba{\begin{eqnarray} }
\def\ea{\end{eqnarray} }

\usepackage{subfig}
\usepackage{bbm}
\usepackage{epsf}
\usepackage{amsmath}
\usepackage{graphicx}
 \usepackage{soul,ulem}
 \usepackage{marginnote}
 
\everymath{\displaystyle}
\usepackage{bm}
\newcommand{\bk}{{\bf k}}
\newcommand{\bq}{{\bf q}}
\newcommand{\bp}{{\bf p}}
\newcommand{\Cia}{{\cal I}^{(a)}}

 \def\calP{{\cal P}}
 
 \def\CalK{{\cal K}}
\usepackage{relsize}

\begin{document}
\title{  Temperature Dependence of   Gluon  and Ghost Propagators  in a   Dyson-Schwinger Equations  context}
 \author{L.~P. Kaptari}
 \affiliation{Bogoliubov Lab.~Theor.~Phys., 141980, JINR, Dubna, Russia}

\begin{abstract}
  We investigate   the    finite-temperature structure of ghost and gluon propagators    within an approach based on the rainbow truncated Dyson-Schwinger equations  in Landau gauge.
 The method, early used for modeling the quark, ghost and gluon propagators in vacuum,
  is extended  to finite temperatures.
  In Euclidean space, within the Matsubara imaginary-time formalism it is necessary to distinguish between
  the transversal and  longitudinal, with respect to the heat bath, gluon dressing functions, for which
  the   Dyson-Schwinger equation  splits  into a corresponding system of  coupled equations. This  system  is considered within the rainbow approximation  generalized   to finite temperatures   and solved numerically. The solutions for the ghost and gluon propagators are obtained  as  functions of temperature $T$, Matsubara frequency  $\Omega_n$ and three-momentum squared ${\bf k}^2$.  The effective parameters of the approach are taken from  our previous fit of the corresponding Dyson-Schwinger solution  to the lattice QCD  data at zero temperature.  In solving the coupled system of the Dyson-Schwinger equations at finite temperatures, the model parameters
  are  treated as constants, independent on temperature.
  It is  found that, for zero Matsubara frequency, the dependence  of the ghost and gluon
   dressing functions on ${\bf k}^2$ are not sensitive to the temperature $T$,  while   at ${\bf k}^2=0$  their dependence on $T$   is quite strong. Dependence on  the Matsubara frequency $\Omega_n$ is investigated as well. The performed numerical analysis of the solution of the Dyson-Schwinger equations
   shows that at certain value of the temperature $T_0\sim 150$ MeV the iteration procedure  does not longer
    converge. In the vicinity of $T_0$ the longitudinal gluon propagator   increases quite fastly, whereas the transversal propagator does not exhibit any irregularity.  This in a qualitative agreement with results obtained within the QCD lattice calculations in this temperature interval.
 \end{abstract}

\maketitle

\section{Introduction}\label{intro}
The  study of the behaviour of hadrons in hot and dense nuclear matter
is among the most interesting and challenging problems intensively investigated by theorists and experimentalists.
It is tightly connected to studies of the quark-gluon plasma (QGP) which is the deconfined  phase of the strongly interacting QCD. It is commonly adopted that this phase existed in the early universe and can also be  briefly
 produced in laboratory by heavy-ion collisions. This stimulates intensive studies of high energy processes with
  heavy ions.  A bulk of the running and projected   experiments  in various research centers, e.g. at
 Belle (Japan),  BESIII (Beijing, China),
 LHC (CERN), GlueX (JLAB, USA), NICA (Dubna, Russia), HIAF (China),  FAIR (GSI, Germany) etc.,
include  in their research programmes comprehensive investigations of properties of hadrons
 at high temperatures  and  the possible  transition  to quark-gluon plasma.
  Appropriate approaches in theoretical study of properties of quarks and gluons  at zero and finite
  temperature are the   lattice QCD simulations~\cite{GhostLatticeMishaPRD,BornyakovLattice,Aouane:2011fv,Ilgenfritz:2017kkp,LatticeQuenchedvsUnquenched,QuenchedUnquenchedGluon,Albanese,Chen,Morningstar}, complemented by functional renormalization group (FRG) methods (see e.g. Refs.~\cite{Dupuis:2020fhh,PawlowskyFRG}), approaches based on QCD Sum Rules~\cite{Shifman:1978bx,Shuryak:1982dp} and the functional approaches via  Dyson-Schwinger equations~\cite{FisherTempQCD} (for a review of different methods
   in studying gauge bosons at zero and finite temperatures, see Refs.~\cite{MaasPhysReport,Das}).
   Despite a fairly rigorous  theoretical
   foundation, these approaches are rather complex and cumbersome
   in further applications in describing the temperature dependence of physical
    bound states, such as mesons and glueballs, for example. It is tempting therefore,
     to elaborate transparent models which, on the one hand,   are simple
     and physically understandable, on the other hand,   covering the main characteristics
     of the studied phenomena. In this lpaper, we employ a generalization of the well known
     rainbow approximation to the truncated Dyson-Schwinger equations (tDSE)~\cite{alkof,rober,MT,Dorkin:2014lxa}
     to finite temperatures in Euclidean space within the imaginary-time
     formalism~\cite{S-XQINprd84,Dorkin:2014lxa,BlankKrass,Viebakh,dor}.

The paper is organized as follows. In Sec.~\ref{sec2} we formulate the truncated  system of  Dyson-Schwinger  equations for the ghost and gluon propagators at finite temperature within the Matsubara imaginary time formalism. The general expressions for the system of
tDSE's   for ghost, transversal and longitudinal gluon propagators are presented. In Sec.~\ref{bow} the rainbow approximation to the interaction   kernels of the integral equations is defined. Further, in  subsections ~\ref{subghost} and \ref{subgluon} this approximation is applied to perform explicitly the angular integrations in the ghost and gluon tDSE and to reduce the four-dimensional integral equations
to   summations over the Matsubara frequencies and one-dimensional integrations over the corresponding loop momenta.
Results of numerical solutions of tDSE are presented in Sec.~\ref{results}. The solution for the ghost dressing
function and transversal and longitudinal gluon propagators are discussed in subsections~\ref{resGhost} and \ref{resGluont}, where the temperature and momentum dependence as well as the dependence on the Matsubara frequency, are analysed in some details. We summarize our conclusions in Sec.~\ref{summary}.

\section{Coupled Dyson-Schwinger equations for gluons and ghosts}\label{sec2}
\label{Dys}
\begin{figure}[!ht]
 \begin{center}
 \includegraphics[scale=0.55 ,angle=0]{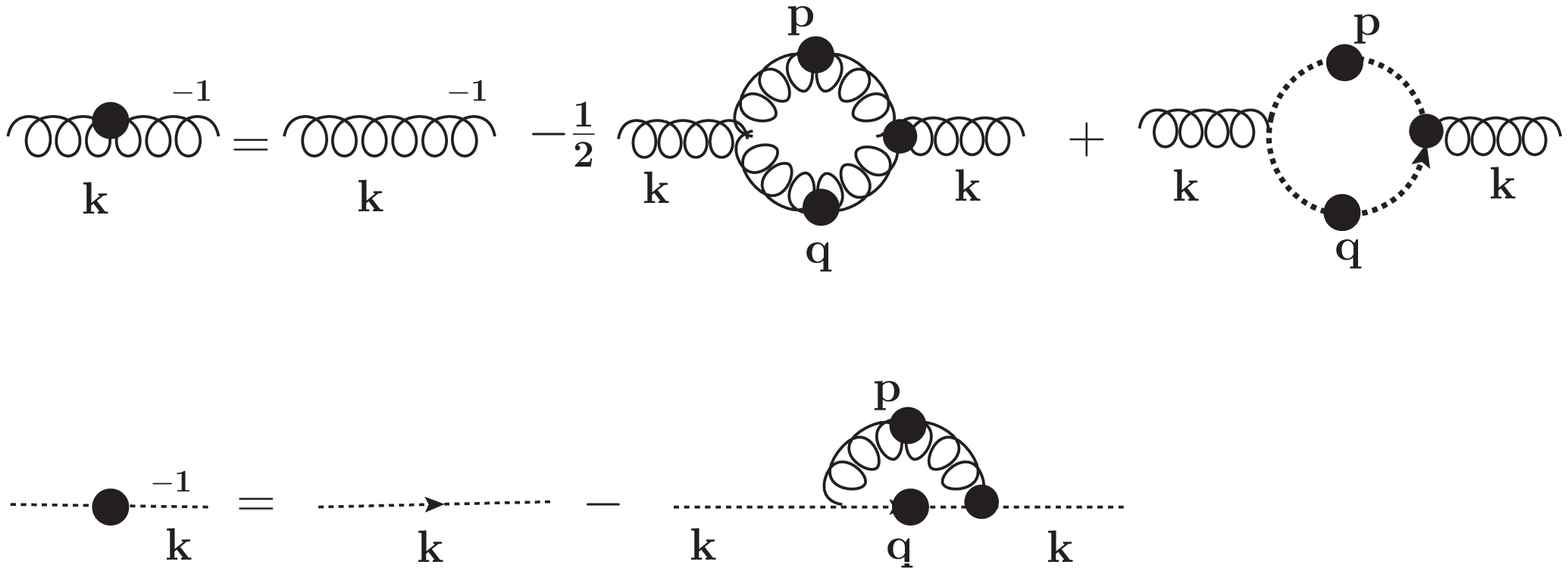}\vspace*{3mm}
 \end{center}
 \caption{ Diagrammatic representation  of the employed Dyson-Schwinger equations for ghost (top line) and
 gluon (bottom line) propagators. The internal wiggly and  dashed  lines denote
   the full,  marked by filled blobs, gluon and ghost propagators, respectively. The irreducible one-particle
   vertices are also denoted by filled blobs. In the gluon Dyson-Schwinger equation,
    terms with four-gluon vertices and quark loops  have been dismissed.}
 \label{dse}
 \end{figure}

A nonperturbative continuum approach to QCD is provided
by the Dyson-Schwinger equations.
The system of Dyson-Schwinger coupled equations for the quark, ghost and   gluon propagators and the vertex functions as well,   being the equations of motion for the corresponding Green's functions, can be
considered as an alternative  integral formulation of QCD. Evidently, attempts to solve exactly the system of these   equations, which relate nonperturbatively  $n$-point and
$n+1$-point functions, thus constituting an infinite  tower of coupled equations,
fail to find  adequate numerical solutions.
 Consequently, for  practical purposes some approximations are necessary.  Usually one employs truncations of the
 whole system by restricting oneself to only few first Feynman diagrams. Typically, the truncations contain only
 one-loop  diagrams with dressed propagators and vertices, as
depicted in Fig.~\ref{dse} (cf., e.g. Ref.~\cite{smekalAnnals267}). The resulting system is referred to as
 the truncated Dyson-Schwinger equation (tDSE).
 Note that this subset is already strongly
reduced relative to the full set of equivalent QCD diagrams, discussed in e.g.~\cite{FisherTempQCD} in a wider context and even relative to the subset of one-loop diagrams.
It does contain neither loops with four-gluon vertices nor quark loops.
The four-gluon vertices are the momentum independent tadpole-like terms and terms with explicit two-loop contributions. While the former leads an irrelevant constant which vanishes perturbatively in Landau gauge, the latter are subleading in the infrared, cf. Ref.~\cite{smekalAnnals267}.
 As for  ignoring   quark loops, it can be justified by the
 observation~\cite{LatticeQuenchedvsUnquenched,QuenchedUnquenchedGluon,fisher} that
in the tDSE the unquenching  effects are rather small for
 the dynamical quark masses. For  the gluon dressing function, such effects  are seen only
 in the neighbourhood of its maximum at $k\sim 0.85-1.0 \, {\rm  GeV/c}$, where
  the  screening effect from the  creation of quark-antiquark pairs from the vacuum
  slightly decreases  the value of the gluon dressing. In our approach, this
  effect is implicitly taken into account by adjusting the phenomenological parameters at zero temperature
  to full, unquenched  lattice calculations, see e.g. Refs.~~\cite{GhostLatticeMishaPRD,BornyakovLattice}
  for SU(2) lattice QCD data.

The  system of coupled tDS equations for the gluon and ghost propagators
is obtained by direct calculation of the corresponding one-loop Feynman diagrams in Fig.~\ref{dse}. In Landau gauge, the resulting system can be written as
\begin{eqnarray} &&
  D_{\mu\nu}^{-1}(k_4^2,\bk^2)=Z_3  D_{0\, \mu\nu}^{ -1}(k^2)   + \Sigma^{gluon}_{\mu \nu}(k_4^2,\bk^2) +
 \Sigma^{ghost}_{\mu \nu}(k_4^2,\bk^2)
\label{sdeGluon},\\[3mm]
&& D_G^{-1}(k_4^2,\bk^2) = \tilde Z_3  D_{0G}^{ -1}(k^2) +S^{ghost} (k_4^2,\bk^2).
\label{sdeGhost}
\end{eqnarray}
 where $Z_3$ and $\tilde Z_3$ are the gluon and ghost renormalization  constants, respectively,
 $D_{0\, \mu\nu}(k^2)$ and  $D_{0G}(k^2)$ denote the corresponding free propagators,  $k^2=k_4^2+\bk^2$ and
 the self-energy terms $\Sigma^{gluon}_{\mu \nu}(k_4,\bk)$,
 $ \Sigma^{ghost}_{\mu \nu}(k_4,\bk)$ and
 $S^{ghost}(k_4,\bk)$ correspond to the three loop diagrams depicted in Fig.~\ref{dse}.
In Euclidean space, at zero temperatures the gluon and ghost propagators are
\begin{equation}
D_{\mu\nu}^{ab}(k^2) =
\delta^{ab} D_{\mu\nu}(k^2)=\delta^{ab}\left( \delta_{\mu\nu} - \frac{k_\mu k_\nu}{k^2}\right)\frac{Z(k^2)}{k^2} ;
\quad  D_G^{ab}(k^2) =\delta^{ab}D_G(k^2)=- \delta^{ab}\frac{G(k^2)}{k^2},
\end{equation}
where $a,b$ are the color indices and $Z(k^2)$ and $G(k^2)$  are the gluon and ghost dressing functions, respectively.

 At finite temperatures,     within the imaginary time formalism   all  the relevant
 four-momenta in Euclidean space possess  discrete  spectra of the corresponding fourth component, $k_4=\Omega_n$ ($n \subset\mathbb{Z}$),  where
$\Omega_n=2\pi n T $ for bosons (gluons) and $\Omega_n= \pi (2n+1) T $ for fermions (quarks).
As for  the Fadeev-Popov ghosts, in spite of the  Grassmann nature of their fields needed in the gauge fixing procedure,
   formally  ghosts are related to (nonexisting) spin-zero particles and to  scalar propagators. Consequently,
 the ghost fields   satisfy periodic conditions like any bosonic field, cf. Refs.~\cite{Das,Gruter,Bernard,Gross},
 so that
 the Mtzubara frequencies for ghosts   are  even
   \footnote{There has been some confusion in the literature about the ghost Matsubara frequency,
   see Ref.~\cite{Baluni}.}.

  In  Landau gauge,  the Euclidean gluon propagator ${\cal D}_{\mu\nu}^{ab}(k_4,\bk)$ and
  ghost propagator ${\cal D}_G^{ab}(k_4,\bk)$ are expressed via the dressing functions $Z(k_4,\bk)$ and $G(k_4,\bk)$  as
 \begin{eqnarray}&&
 {\cal D}_{\mu\nu}^{ab}(k) =  \delta^{ab}{ D}_{\mu\nu}(k_4,\bk)= \delta^{ab}
 \left\{\frac{Z_T(k_4,\bk)}{k^2}\calP^T_{\mu\nu}(\bk)+\frac{Z_L(k_4,\bk)}{k^2}\calP^L_{\mu\nu}(k_4,\bk)
 \right\}, \label{propgluon} \\ &&
 {\cal D}_G^{ab}(k_4,\bk)= - \delta^{ab} D_G(k_4,\bk)= -\delta^{ab}\frac{G( k_4,\bk )}{k^2},
 \label{propghost}
 \end{eqnarray}
 where $\calP^L_{\mu\nu}(k_4,\bk)$, and   $\calP ^T_{\mu\nu}(\bk)$ are
  the longitudinal and transversal, in 3D space, projectors,
\begin{eqnarray}&&
\calP ^T_{\mu\nu}(\bk)=\left\{  \begin{array}{llll}
                                         0,&& \mu \ {\rm and/or} \  \nu=4,  \\
                                         \delta_{\alpha\beta}-\dfrac{k_\alpha k_\beta}{\bk^2};&&\mu,\nu=\alpha,\beta=1,2,3,
                                         \end{array}\right.
     \label{trans}\\[1mm] &&
\calP ^L_{\mu\nu}(k) =\calP_{\mu\nu}(k)-\calP ^T_{\mu\nu}(\bk); \quad {\rm where}\quad \calP_{\mu\nu}(k)=\delta_{\mu\nu}-\dfrac{k_\mu k_\nu}{k^2}.
\label{proj}
\end{eqnarray}
Note that these  projection operators are four-dimensionally transverse and have the properties
\begin{equation}
\calP ^{T} \calP ^{T}=\calP ^{T},\,\calP ^{L} \calP ^{L}=\calP ^{L},\,
\calP ^{T} \calP ^{L}=\calP ^{L} \calP ^{T}=0 , \quad
\calP ^{T}_{\mu\nu} \calP ^{T}_{\mu\nu} =2, \,  \calP ^{L}_{\mu\nu} \calP ^{L}_{\mu\nu}  =1.
\label{relations}
\end{equation}
 Relations (\ref{relations}) are widely  used in deriving the Dyson-Schwinger equations for transverse and longitudinal
dressing functions  $Z_{(T,L)}$ from eqs.~(\ref{sdeGluon})-(\ref{propgluon}).
It should be noted that, in an arbitrary gauge  the gluon propagator
reads as
 \begin{eqnarray}&&
 {\cal D}_{\mu\nu} (k) =
  \frac{Z_T(k_4,\bk)}{k^2} \calP^T_{\mu\nu}(\bk)+ \frac{Z_L(k_4,\bk)}{k^2}\calP^L_{\mu\nu}(k_4,\bk)
  + \frac{\eta}{k^2} \frac{k_\mu  k\nu}{k^2},
 \label{propgluon1}
 \end{eqnarray}
 where $\eta$ is the gauge parameter. The inverse of (\ref{propgluon1} ) is
 \begin{eqnarray}&&
 {\cal D}_{\mu\nu}^{-1}(k) =
 \frac{k^2}{Z_T(k_4,\bk)}\calP^T_{\mu\nu}(\bk)+\frac{k^2}{Z_L(k_4,\bk)}\calP^L_{\mu\nu}(k_4,\bk)
 +  \frac{k^2}{\eta} \frac{k_\mu k_\nu}{k^2}  \label{propgluonInverse}.
 \end{eqnarray}

 It can be seen that in Landau gauge ($\eta=0$) the inverse gluon propagator becomes ill-defined. However, since
 the longitudinal part of the propagator $\eta \frac{k_\mu  k\nu}{k^2}$ is not affected by the renormalization procedure,
  in Eq.~(\ref{sdeGluon})
 the terms with $\eta^{-1}$ in the left hand side and right hand side cancel each other and the resulting equation
 remains well defined. Generally, performing calculations with propagators in the Landau gauge one shall
use the propagator in the form of Eq.~(\ref{propgluon1}) and take the limit $\eta\to 0$ at the end of calculations.
 In our case the tDSE, Eq.~(\ref{sdeGluon}), does not depend on $\eta$ and, consequently, we can neglect such terms   from the very beginning and write Eq.~(\ref{sdeGluon}) as
 \begin{eqnarray} \label{DSEa}
  \frac{k^2}{Z_T(k_4,\bk)} \calP^T_{\mu\nu}(\bk) +   \frac{k^2}{Z_L(k_4,\bk)} \calP^L_{\mu\nu}(k_4,\bk)=
   Z_3  k^2  \calP_{\mu\nu}(k) +
 \Sigma^{gluon}_{\mu \nu}(k_4^2,\bk^2) +
 \Sigma^{ghost}_{\mu \nu}(k_4^2,\bk^2).
 \end{eqnarray}
   Then, after contracting the color indices in the corresponding loops in Fig.~\ref{dse},
    the self-energy parts  explicitly read as
 \begin{eqnarray}
   \Sigma^{gluon}_{\mu \nu}(\Omega_n,\bk) && = \frac12\frac {3T}{ 2\pi^2 }\sum_m\int d^3 q
   \Gamma_{\mu\alpha\rho}^{(0)}( k,-q,p)
\frac{Z_T(\Omega_m,\bq)  \calP^T_{\alpha\beta}(\bq)+Z_L(\Omega_m,\bq)\calP^L_{\alpha\beta}}{q^2} \bigg[\frac{g^2}{4\pi} \nonumber \times  \\ &&
\Gamma_{\beta\nu\sigma}(q,-p,-k)
\left(D_T(\Omega_{mn},\bp)\calP^T_{\sigma\rho}(\bp) + D_L(\Omega_{mn},\bp)\calP^L_{\sigma\rho}(p)\right)
\bigg],
   \label{selfGluon}\\[3mm]
\mathlarger\Sigma^{ghost}_{\mu \nu}(\Omega_n,\bk)&&=-\frac{3T}{2\pi^2}\sum_m\int d^3q \frac{G(\Omega_m,\bq)}{q^2}
\left [ \Gamma^{(0)}_\mu(q)\frac{g^2}{4\pi}D_G( \Omega_{mn},\bp)\Gamma_\nu(p)\right] ,
\label{selfGhost}\\[3mm]
S^{ghost}(\Omega_n,\bk)&&= \frac {3T}{ 2\pi^2 } \sum_m \int
d^3 q   \left[\frac{g^2}{4\pi} \Gamma_{\mu}^{(0)}(q)
D_{\mu\nu}(p^2) \Gamma_{\nu}(k,q,p) \right]
 \frac{G(\Omega_m,\bq)}{q^2},
 \label{Sghost}
\end{eqnarray}
where $p=q-k$, $p_4\equiv\Omega_{mn}=\Omega_{m}-\Omega_{n}$, $\bp =\bq-\bk  $
and  the terms enclosed in the square brackets define the interaction kernel  of the corresponding integral equation.
In the above expressions we have taken into account that due to periodicity of the gluon and ghost fields, the integration over the fourth component $dq_4$ is replaced by the   summation over the discrete  Matsubara frequencies according to
\begin{equation}
 \int\limits_{-\infty}^\infty \frac{dq_4}{2\pi} \int \frac{d^3 q  }{(2\pi)^3} f(q_4,\bq) \rightarrow
 T\sum\limits_{n=-\infty}^\infty   \int \frac{d^3q  }{(2\pi)^3} f(\Omega_n,\bq).
 \end{equation}
    The tDSE for  the  scalar dressing functions $Z_{T,L}$ are obtained from  the tDSE (\ref{sdeGluon})
    by inserting  Eqs.~(\ref{selfGluon})-(\ref{selfGhost}) in to Eq.~(\ref{sdeGluon}),
 multiplying the left
 and right hand sides consecutively  by $\calP^{T}(\Omega_n,\bk)$ and $\calP^{L}(\Omega_n,\bk)$ and
  contracting all the Lorentz indices.

The gluon renormalization constants $  Z_{3(T,L)}$ for transverse and longitudinal propagators, in principle, can be  different and must be determined separately.
 In the present paper, for the sake of simplicity, we adopt the renormalization constant  $ Z_3$  to be the same
 for both, transverse and longitudinal parts. Moreover, the concrete numerical values  for
 $ Z_3$ and $ \tilde Z_3$ are taken from the previous fit~\cite{ourGlueball} of the tDSE solution in vacuum   to the lattice QCD data, viz. $ Z_3=\tilde Z_3\approx 1.065$.

\section{Rainbow approximation }\label{bow}
Attempts to solve the system of coupled equations by employing directly the QCD Feynman rules
 encounter  difficulties related to regularizations of divergent
loop integrals and to symmetry constraints on gluon-ghost and three-gluon vertices,
such as Slavnov-Taylor identities. Clearly,  further approximations are required.
In vacuum, the simplest approximation consists in  replacements of the fully dressed
three-gluon and  ghost-gluon vertices by their bare values,
known as the Mandelstam approximation~\cite{CPCMandelstam,MANDELSTAM_Approx,PennigtonMandels}
and as the y-max approximation~\cite{YmaxApprox}. Within these approaches, in order  to
 facilitate the angular integrations in  analytical and numerical
analysis of Eqs.~(\ref{sdeGluon}) - (\ref{sdeGhost}),
additional simplifications for the momentum  dependence in $Z(k^2)$ and $G(k^2)$ have been
 adopted.  Also, some ansatzs for infrared and ultraviolet asymptotic solution  $Z(k^2)$ and $G(k^2)$
   have been adopted. A more rigorous analysis of the tDSE has been presented  in a series of
 publications in Refs.~\cite{IRGluonProp_AlkoferSmekal,smekalAnnals267}
 (see also the review Ref.~\cite{AlkoferphysRep} and references therein quoted).
All the above mentioned approaches result in rather cumbersome expressions for
the system of tDSE's which, consequently, cause  difficulties in finding the numerical solutions.
Moreover, these circumstances are more significant in attempts to solve the tDSE at finite temperatures.
 In order to avoid the mentioned difficulties  and to simplify
  the angular integrations,  in the present  paper  we employ a rainbow-like approximation,
similar to the rainbow model~\cite{alkof,rober,BlankKrass,Viebakh,dor}, for the interaction kernels enclosed in square brackets  in Eqs.~(\ref{selfGluon}) - (\ref{Sghost}).
  This   approximation consists in
replacing the dressed vertices together with the dressed exchanging propagators by their bare
quantities augmented by some effective form factors:
\begin{eqnarray}&&
\left[\frac{g^2}{4\pi} \Gamma_{\mu \alpha\rho}^{(0)}(k,-q,p)
D_{\rho\sigma}(p^2)\Gamma_{\beta\nu\sigma}(q,-p,-k)\right] = \nonumber \\ &&
  \Gamma_{\mu\alpha\rho}^{(0)}(k,-q,p)
  \bigg [\calP^T_{\rho\sigma}(\bp)F^{T}_1(p^2)+\calP^L_{\rho\sigma}(p)F^{L}_1(p^2)\bigg]
  \Gamma_{\beta\nu\sigma}^{(0)}(q,-k,-p),
  \label{rain2}\\[3mm]
  &&
 \left [\frac{g^2}{4\pi}\Gamma_{\mu}^{(0)}(q) D_G(p^2)\Gamma_\nu(k)\right]=\Gamma_\mu^{(0)}(q)
 F_2(p^2)\Gamma_\nu^{(0)}(p^2), \label{rain3}
  \\ &&
\bigg[\frac{g^2}{4\pi} \Gamma^{(0)}_{\mu }(q) D_{\mu\nu}(p^2) \Gamma_{\nu}(k,q,p) \bigg]
= \Gamma_{\mu}^{(0)}(q) \left[ \calP^T_{\mu\nu}(\bp)F^{T}_3( p^2)+\calP^L_{\mu\nu}(p)F^{L}_3(p^2)\right]\Gamma_{\nu}^{(0)}(k),
\label{rain1}
\end{eqnarray}
 where the superscript (0)  of the vertex functions $\Gamma$ means the free 3-gluon and free ghost-gluon-ghost
 vertices (see Ref.~\cite{ourGlueball} for more details).
As in Ref.~\cite{ourGlueball}, we
 use for  the form-factors $F_{1-3} (p^2)$  a Gaussian form with two terms for each formafactor $F_{1-3}(p^2)$.
This is quite sufficient to obtain a reliable solution of the system of tDS equations at $T=0$ when the O(4)
symmetry holds exactly and  consequently, the longitudinal and transversal part of the gluon propagator coincide.
 Explicitly, in Euclidean space the effective form factors are chosen as
\begin{eqnarray} &&
F^{T,L}_{1}(p^2)=  D^{T,L}_{11} \frac{  p^2}{  (\omega_{11}   ^{  T,L })^6  }
e^{  - p^2/(\omega_{11}^ {T,L})^2 }+
D^{T,L}_{12} \frac{  p^2}{ ( \omega_{12}  ^{  T,L })^6 }
e^{  - p^2/(\omega_{12}^ {T,L})^2 } ,\label{ff1} \\ &&
F_{2}(p^2)=  \frac{D_{21}}{\omega_{21}^4} e^{ -  p^2/\omega_{21}^2} +
\frac{D_{22}}{\omega_{22}^4} e^{ -  p^2/\omega_{22}^2 }, \label{ff2}
\\ &&
F^{T,L}_{3}(p^2)=  D^{T,L}_{31} \frac{  p^2}{  (\omega_{31} ^{  T,L })^6 }
e^{  - p^2/(\omega_{31}^ {T,L})^2 }+
D^{T,L}_{32 } \frac{  p^2}{  (\omega_{32} ^{  T,L })^6 }
e^{  - p^2/(\omega_{32}^ {T,L})^2 }
 \label{ff3},
\end{eqnarray}
with phenomenological parameters $D_i$ and $\omega_i$  fitted to  provide a
reasonable good   description of the lattice SU(2) data at zero temperatures (cf.~Ref.~\cite{ourGlueball}).

\subsection{The tDSE for the ghost propagator, Fig.~\ref{dse}, bottom line}\label{subghost}
With   the effective interaction eqs.~(\ref{rain2})-(\ref{ff3}),  the angular integration
can be  carried out analytically~\cite{ourGlueball} resulting in a system of linear algebraic equations with respect to the Matsubara frequency and  one-dimensional integral equations with respect to three-momentum $|\bk|$ in Euclidean space.
Explicitly, by using the reltaions
$ \Gamma_\mu^{(0)} (q) = -iq_\mu$ and $ \Gamma_\nu^{(0)} (k) = -ik_\nu$ and
  $q_\mu\, \calP^{(T,L)}_{\mu\nu}(p) k_\nu = q_\mu\calP^{(T,L)}_{\mu\nu}(p)\, q_\nu$ and
  Eqs.~(\ref{sdeGhost}), (\ref{proj}) and (\ref{selfGhost}),  the tDSE for ghosts reads as
\begin{eqnarray}
\frac{1}{G(\Omega_n,{\bf k})}=
  \tilde Z_3 -  \frac{3}{2\pi^2}
 T\sum_m \int d^3 q\frac{G(\Omega_m,{\bf q})}{k^2  q^2}
  \left [ \Delta F_3 (p^2) q_\mu \calP^T_{\mu\nu}({\bf p}) q_\nu +F_3^L (p^2) q_\mu P_{\mu\nu}(p) q_\nu\right ],
\label{GhostDSE1}
\end{eqnarray}
where $q=p+k$, $\Delta F_3 (p^2)=F_3^T (p^2)-F_3^L (p^2)$ and
\begin{equation}
q_\mu \calP^T_{\mu\nu}({\bf p}) q_\nu  = \frac{{\bf q}^2 {\bf k}^2}{{\bf p^2}} \sin^2\theta_{\bf k q};
\quad
q_\mu P_{\mu\nu}(p) k_\nu=q_\mu P_{\mu\nu}(p) q_\nu=\frac{q^2 k^2-(qk)^2}{p^2}.
\label{svertki}
\end{equation}

In the limit $T\to 0$ the $O(4)$ symmetry is restored, i.e. $\Delta F_3 (p^2)=0$,
$T\sum\limits_m \to  \int\limits_{-\infty}^\infty\displaystyle\frac{dq_4}{2\pi}$   and
\begin{eqnarray}\!\!\!\!\!\!
\frac{1}{G(  k^2)} = \tilde Z_3-\sum_{j=1}^2 \frac{3}{ \pi^2  }\int   q ^3 d q
    \frac{ G(  q^2)D_{3j}^L}{(\omega_{3j}^L)^6} \exp\left(- (k^2+q^2-2 kq\cos \chi_{kq})/(\omega_{3j}^L)^2\right)\sin^4\chi_{ q  k } d\chi_{ q k} ,
\end{eqnarray}
where  $\chi_{ q k} $ is the hyper-angle between the 4-momenta $  k$ and $ q$.

 By taking into account  that
\begin{equation}
\int  \exp\left[ 2 kq\cos \chi_{kq})/(\omega_{3j}^L)^2\right])\sin^4\chi_{ q  k } d\chi_{ q k} =
\frac{3\pi}{4}\frac{(\omega_{3j}^L)^4}{k^2q^2}
  I_2\left(\frac{2k q}{(\omega_{3j}^L)^2}\right),
  \label{old}
\end{equation}
 where $I_2\left(\frac{2k q}{\omega_{3j}^L)^2}\right)$ is the modified Bessel function,
 the ghost DSE in vacuum is recovered~\cite{ourEPJP}.

 Coming back to finite temperatures, Eq.~(\ref{GhostDSE1}), the DSE for ghosts  within rainbow approximation (\ref{ff3})
reads as
\begin{eqnarray} &&
\frac{1}{G(\Omega_n,\bk)} = \tilde Z_3-
\frac{3T}{2\pi^2} \sum_{m=-\infty}^\infty \sum_{j=1}^2\int d\bq\frac{G(\Omega_m,\bq)}{k^2 q^2}
\bigg\{ {\rm e}^{-p^2/(\omega_{3j}^L)^2}\frac{D^L_{3j}}{(\omega_{3j}^L)^6} \big[
k^2 q^2-(k q)^2\big] +\nonumber \\ &&
\bk^2 \bq^2\sin^2\theta_{\bq \bk} \bigg [1+\frac{\Omega_{mn}^2}{\bp^2}\bigg ]
\bigg[ \frac{D^T_{3j}}{(\omega_{3j}^T)^6}{\rm e}^{-p^2/(\omega_{3j}^T)^2}-
\frac{D^L_{3j}}{(\omega_{3j}^L)^6}{\rm e}^{-p^2/(\omega_{3j}^L)^2}\bigg ]\bigg\},
\label{ghostDSE2}
\end{eqnarray}
where $p^2=(\Omega_n-\Omega_m)^2+\bq^2+\bk^2 - 2|\bk||\bq|\cos\theta_{\bq \bk}\,$, with $\theta_{\bq \bk}$ as the angle between the vectors $\bq$ and $\bk$. Expression
(\ref{ghostDSE2}) permits to carry out angular integration
  $d\bq=2\pi |\bq|^2 d|\bq| \sin\theta_{\bq\bk} d \theta_{\bq\bk} $
 over $\xi=\cos\theta_{\bq\bk}$ explicitly.
  The results is

  \begin{eqnarray} && \frac{1}{G(\Omega_n, \bk)}=
\tilde Z_3-\frac{3T}{ \pi} \sum_{m=-\infty}^\infty   \sum_{j=1}^2
\int |\bq |^2 d |\bq|  \frac{G(\Omega_m,\bq)}{k^2 q^2} \times\nonumber
 \\ && \left\{
\frac{D_{3j}^L}{(\omega_{3j}^L)^6}{\rm e}^{-\Omega_{mn}^2/(\omega_{3j}^L)^2}
\bigg[-|\bk|^2 |\bq|^2\Cia_2(\alpha_L)-2 |\bk||\bq|\Omega_n\Omega_m \Cia_1(\alpha_L)+
\right .
\nonumber\\ &&
\left( |\bk|^2 |\bq|^2+|\bk|^2\Omega_m^2+|\bq|^2\Omega_n^2\right) \Cia_0(\alpha_L) \bigg]
   {\rm e}^{-(|\bq|-|\bk|)^2/(\omega_{3j}^L)^2}
  +\nonumber\\ &&
 \frac{D_{3j}^T}{(\omega_{3j}^T)^6}{\rm e}^{-\Omega_{mn}^2/(\omega_{3j}^T)^2}
 \left( \frac{|\bk| |\bq|\Omega_{mn}^2}{2}\Delta\CalK_b(\alpha_T,b)
 -|\bk|  |\bq| (\omega_{3j}^T)^2 {\rm e}^{-(|\bq|-|\bk|)^2/(\omega_{3j}^T)^2} \Cia_1(\alpha_T)\right)
-\nonumber \\ && \left .
  \frac{D_{3j}^L}{(\omega_{3j}^L)^6}{\rm e}^{-\Omega_{mn}^2/(\omega_{3j}^L)^2}\left (
  \frac{|\bk||\bq| \Omega_{mn}^2}{2}\Delta\CalK_b(\alpha_L,b)
 -|\bk|  |\bq| (\omega_{3j}^L)^2  {\rm e}^{-(|\bq|-|\bk|)^2/(\omega_{3j}^L)^2} \Cia_1(\alpha_L)
 \right)
  \right\},
  \label{DSE2}
   \end{eqnarray}
  where we introduced the following notation
 \begin{eqnarray}  &&
\Cia_n(\alpha) = {\rm e}^{-\alpha} \int\limits_{-1}^1  {\rm e}^{\alpha \xi} \xi^n d \xi  ; \quad
\CalK_n(\alpha,b) = {\rm e}^{-\alpha b}  \int\limits_{-1}^1 d\xi {\rm e}^{\alpha \xi}\frac{\xi^n }{(b- \xi)} d \xi;
\label{IntaAndIntb} \\ &&
\Delta\CalK_b(\alpha,b)=\CalK_2(\alpha,b)-\CalK_0(\alpha,b);\quad
\alpha_{(L,T)}=\frac{2|\bq| |\bk|}{\left(\omega_{3j}^{(L,T)}\right)^2};
 \quad b=\frac{|\bq|^2+ |\bk|^2}{2|\bq| |\bk|}.
 \label{IntaDeltaK}
\end{eqnarray}
   In particular
\begin{eqnarray} &&
\Cia_0(\alpha) = 2{\rm e}^{-\alpha} \frac{\sinh(\alpha)}{\alpha}; \quad
\Cia_1(\alpha) = \frac1\alpha\bigg[ 2 {\rm e}^{-\alpha} \cosh(\alpha) - \Cia_0(\alpha)\bigg] ;  \nonumber \\ &&
\CalK_0(\alpha,b)=Ei^{(1)}(\alpha (b-1))-Ei^{(1)}(\alpha (b+1));
\CalK_1(\alpha,b)=b\, \CalK_0(\alpha,b) -  {\rm e}^{-\alpha (b-1)}  \Cia_0(\alpha);\nonumber \\ &&
\Cia_2(\alpha)=\Cia_0(\alpha)-\frac2a \Cia_1(\alpha);\, \CalK_2(\alpha,b)=b\, \CalK_1(\alpha,b) +  {\rm e}^{-\alpha ( b-1)}\Cia_1(\alpha).
\end{eqnarray}
where $Ei^{(1)}(x)$ is the exponential integral.
Note that
$\Cia_0(\alpha)$ and $\Cia_1(\alpha)$ are directly related to the modified cylindrical Bessel functions:
\begin{equation}
 \Cia_0(\alpha) =2 {\rm e}^{-\alpha} \sqrt{\frac{\pi}{2\alpha}} I_{\frac12}(\alpha);
 \quad \Cia_1(\alpha) =2 {\rm e}^{-\alpha} \sqrt{\frac{\pi}{2\alpha}} I_{\frac32}(\alpha).
\end{equation}
\subsection{The tDSE for the gluon propagators}\label{subgluon}
The system of coupled equations for the transverse, $T$, and longitudinal,  $L$,  gluon propagators
 is obtained by multiplying Eqs.~(\ref{sdeGluon}) consecutively by the projections operators
$\calP^T_{\mu\nu}(\bk)$ and $\calP ^L_{\mu\nu}(k)$, Eqs.~(\ref{trans})-(\ref{proj}), with  subsequent  contraction of the Lorenz indices $\mu$ and $\nu$.
 \begin{eqnarray} &&
  \frac{2k^2}{Z_T(k_4,\bk)}   =
 2k^2  Z_3      +
 \calP^T_{\mu\nu}(\bk) \Sigma^{gluon}_{\mu \nu}(k_4^2,\bk^2) +
 \calP^T_{\mu\nu}(\bk) \Sigma^{ghost}_{\mu \nu}(k_4^2,\bk^2),  \label{DSET} \\ &&
 \frac{k^2}{Z_L(k_4,\bk)}   =
  k^2 Z_3      +
 \calP^L_{\mu\nu}( k) \Sigma^{gluon}_{\mu \nu}(k_4^2,\bk^2) +
 \calP^L_{\mu\nu}( k) \Sigma^{ghost}_{\mu \nu}(k_4^2,\bk^2) \label{DSEL}.
 \end{eqnarray}
As already mentioned,  the renormalization constant $Z_3$ generally can be different for the transversal and longitudinal
dressing functions. However, for the sake of simplicity, in the present paper we consider it to be the same in both
equations, (\ref{DSET}) and  (\ref{DSEL}).
\subsubsection{The ghost self-energy term $\Sigma^{ghost}_{\mu\nu}$}
 The Lorentz structure of $\Sigma^{ghost}_{\mu\nu}$ is determined by the free ghost-gluon-ghost vertices, see Eq.~(\ref{selfGhost}), which in Euclidean space are $\Gamma^{(0)}_{\mu }(q)=-iq_\mu$ and $\Gamma^{(0)}_{\nu }(q)=-ip_\nu$.
Since $k_\nu \calP^{T,L}_{\mu\nu}(\bk) =0$  the Lorentz structure of $\mathlarger\Sigma^{ghost}_{\mu\nu}$ is $\sim q_\mu q_\nu$ and the relevant Lorentz contractions are
\ba
q_\mu\calP^T_{\mu\nu}(\bk)  q_\nu=\bq^2\sin^2\theta_{\bk \bq} ;\qquad
q_\mu\calP_{\mu\nu}(k)  q_\nu=\frac{1}{k^2} \left [ k^2 q^2 - (k q)^2\right],
\label{TmunuSigma}
\ea
 where $k^2 = k_4^2 + \bk^2$ and $q^2 = q_4^2 + \bq^2$.
The longitudinal part is defined by

 \begin{equation}
 \calP^L_{\mu\nu}( k)\mathlarger\Sigma^{ghost}_{\mu\nu}(\Omega_n,\bk)=
 \calP_{\mu\nu}(\bk)\mathlarger\Sigma^{ghost}_{\mu\nu}(\Omega_n,\bk) - \calP^T_{\mu\nu}(\bk)\mathlarger\Sigma^{ghost}_{\mu\nu}(\Omega_n,\bk).
 \label{PmunuLGhost}
 \end{equation}
Direct calculations of the transversal and longitudinal parts provide
 \begin{eqnarray} &&
 \calP^T_{\mu\nu}(\bk)\mathlarger\Sigma^{ghost}_{\mu\nu}(\Omega_n,\bk)=
  \frac{3T}{|\bk|\pi}\sum_{i=1}^2 \frac{D_{2i}}{\left(\omega_{2i}^T\right)^2 }\sum_{m=-\infty}^\infty
 \int d|\bq|  \frac{\bq^3}{q^2} G(q^2){\rm e}^{-(\Omega_{mn}^2 + (|\bk|-|\bq|)^2)/\left(\omega_{2i}^T\right)^2}
 \Cia_1(\alpha_i^T), \nonumber \\ \label{TmunuGhost}  \\ &&
 \calP_{\mu\nu}(k)\mathlarger\Sigma^{ghost}_{\mu\nu}(\Omega_n,\bk)=
  \frac{3T}{\pi k^2}\sum_{i=1}^2 \frac{D_{2i}}{\left(\omega_{2i}^L\right)^4 }
  \sum_{m=-\infty}^\infty \times\nonumber \\[1mm] &&
 \int d|\bq|  \frac{\bq^2}{q^2} G(q^2){\rm e}^{-(\Omega_{mn}^2 + (|\bk|-|\bq|)^2)/\left(\omega_{2i}^L\right)^2}
 \int_{-1}^1 d\xi {\rm e}^{\alpha_i^L (\xi-1)} \big [
 k^2 q^2 -(k q)^2\big] , \label{PmunuGhost}
 \ea

 where the angular integration in the last equation is
 \ba  &&
  \int_{-1}^1 d\xi {\rm e}^{\alpha_i^L (\xi-1)} \big [k^2 q^2 -(k q)^2\big] = 
  \bigg[\Omega_m^2 |\bk|^2 + \Omega_n^2 |\bq|^2\bigg]\Cia_0(\alpha_i^L)+
 |\bq|  |\bk| \left[ \left(\omega_{2i}^L\right)^2  -2  \Omega_m \Omega_n \right]\Cia_1(\alpha_i^L).\nonumber
 \\ &&
 \label{integrals} \ea
In the above expressions  $\alpha_i^{L,T} = \displaystyle\frac{2|\bk| |\bq|}{\left(\omega_{2i}^{L,T}\right)^2}$.

\subsubsection{The gluon self-energy term $\Sigma^{gluon}_{\mu\nu}$.}

Calculations of the gluon self-energy part $\mathlarger\Sigma^{gluon}_{\mu \nu}(\Omega_n,\bk)$ are much more
involved. Within the rainbow approximation one has
\ba &&
  \Sigma^{gluon}_{\mu \nu}(\Omega_n,\bk)  =\frac12 \frac{3T}{2\pi^2}\sum_m \int d^3\bq\,
  \Gamma_{\mu\alpha\rho}^{(0)}(p,k,-q) \Gamma_{\beta\nu\sigma}^{(0)}(q,-p,-k) \nonumber \\ &&
  \frac{1}{q^2} \bigg [
   \Delta F_1^{T,L}(p_4,\bp)\calP_{\rho\sigma}^T(\bp)  + F_1^L(p_4,\bp) \calP_{\rho\sigma}(p)\bigg]
   \bigg[ \Delta Z(\Omega_m,\bq) \calP_{\alpha\beta}^T(\bq) + Z_L(\Omega_m,\bq) \calP_{\alpha\beta}(q)\bigg]
\ea
where
$\Delta Z(\Omega_m,\bq) = Z_T(\Omega_m,\bq) -Z_L(\Omega_m,\bq)$
 and $\Delta  F_1(p^2)=F_1^T(p^2)-F_1^L(p^2)$ with $p=(p_4,\bp) =(\Omega_m-\Omega_n, \bq-\bk)$.
The
 "reduced" vertices $\Gamma_{\mu\alpha\rho}^{(0)}(p,k,-q)$ and $ \Gamma_{\beta\nu\sigma}^{(0)}(q,-p,-k)$ defined via the
free three-gluon vertices  as
  $G^{abc}_{\mu\alpha\rho} \equiv f^{abc}\Gamma^{(0)}_{\mu\alpha\rho}$
  and $G^{bac}_{\beta\nu\sigma} \equiv f^{bac}\Gamma^{(0)}_{\beta\nu\sigma}$     are symbolically presented in Fig.~\ref{G1G2}.
\begin{figure}[!h]
 \includegraphics[scale=0.42 ,angle=0]{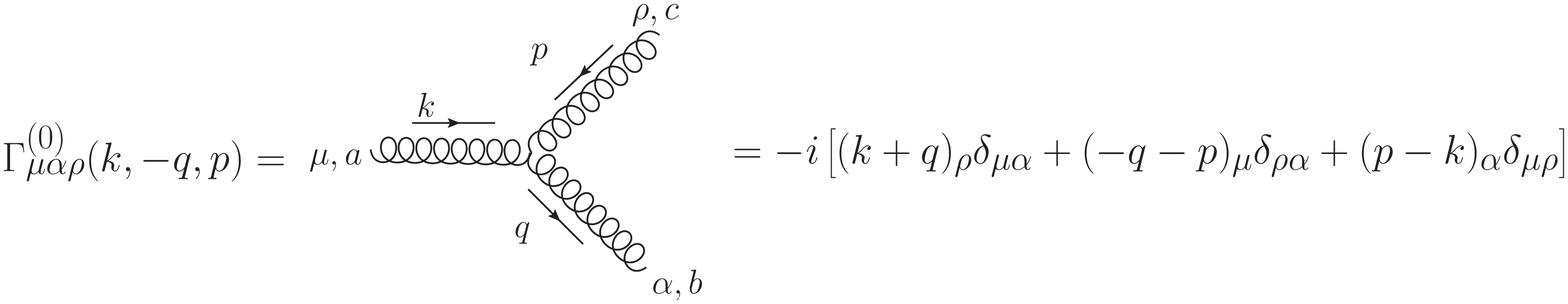}\vspace*{3mm}
 \includegraphics[scale=0.39 ,angle=0]{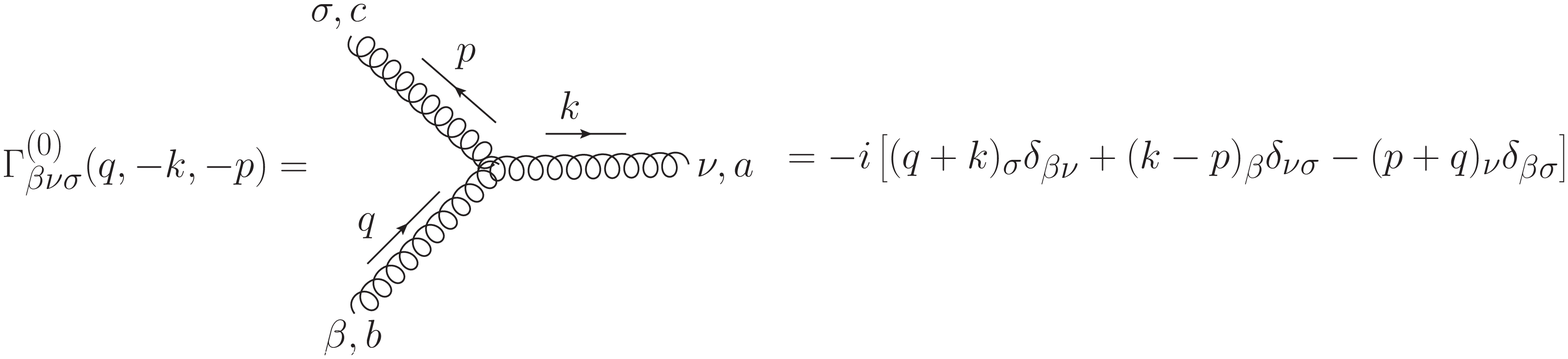}\vspace*{3mm}
 \caption{Diagrammatic representation of the free   three-gluon vertices $ \Gamma^{(0)}_{\mu\alpha\rho}$
  and $ \Gamma^{(0)}_{\beta\nu\sigma}$.   The  Lorentz and color indices
   associated with each gluon momentum $k,q,p$, are  denoted as  $\mu,\nu,\alpha,\beta,\rho,\sigma$ and
  $a,b,c$, respectively. The corresponding color factors $f^{abc}$ and $f^{bac}$,
    accompanying each vertex have been already accounted for in  Eqs.~(\ref{selfGluon})-(\ref{Sghost}) through
    the identity $f^{abc} f^{bac}=-3$.}
 \label{G1G2}
  \end{figure}
With these quantities, the system of tDSE coupled equations for the transversal and longitudinal dressing functions are
\ba &&
\frac{1}{Z_T} =Z_3   + \frac{1}{2k^2}
\frac{3T}{|\bk|\pi}\sum_{i=1}^2 \frac{D_{2i}}{\left(\omega_{2i}^T\right)^2 }\sum_{m=-\infty}^\infty
 \int d|\bq|  \frac{\bq^3}{q^2} G(q^2){\rm e}^{-(\Omega_{mn}^2 + (|\bk|-|\bq|)^2)/\left(\omega_{2i}^T\right)^2}
 \Cia_1(\alpha_i^T) +\nonumber \\ &&
\frac{1}{2k^2}
  \frac{3T}{2\pi}\sum_m \int d|\bq| |\bq|^2 \int_{-1}^1 d\xi\, \bigg\{  \calP^T_{\mu\nu}(\bk)
  \Gamma_{\mu\alpha\rho}^{(0)}(p,k,-q) \Gamma_{\beta\nu\sigma}^{(0)}(q,-p,-k) \times\label{DSETrans} \\ &&
  \frac{1}{q^2} \bigg [
   \Delta F_1^{T,L}(p_4,\bp)\calP_{\rho\sigma}^T(\bp)  + F_1^L(p_4,\bp) \calP_{\rho\sigma}(p)\bigg]
   \bigg[ \Delta Z(\Omega_m,\bq) \calP_{\alpha\beta}^T(\bq) + Z_L(\Omega_m,\bq) \calP_{\alpha\beta}(q)\bigg]\bigg\}
\nonumber
\ea
for the transversal dressing $Z_T(k^2)$ and
\ba &&
\frac{1}{Z_L} =Z_3   +\frac{1}{k^2}[ \calP_{\mu\nu}^L(k)]\Sigma^{ghost}_{\mu\nu}(\Omega_n,\bk) +\label{DSELong} \\ &&
\frac{1}{k^2}
  \frac{3T}{2\pi}\sum_m \int d|\bq| |\bq|^2 \int_{-1}^1 d\xi\, \bigg\{   \calP_{\mu\nu}^L(k)
  \Gamma_{\mu\alpha\rho}^{(0)}(p,k,-q) \Gamma_{\beta\nu\sigma}(q,-p,-k) \times\nonumber \\ &&
  \frac{1}{q^2} \bigg [
   \Delta F_1^{T,L}(p_4,\bp)\calP_{\rho\sigma}^T(\bp)  + F_1^L(p_4,\bp) \calP_{\rho\sigma}(p)\bigg]
   \bigg[ \Delta Z(\Omega_m,\bq) \calP_{\alpha\beta}^T(\bq) + Z_L(\Omega_m,\bq) \calP_{\alpha\beta}(q)\bigg]\bigg\},
\nonumber
\ea
for the longitudinal $Z_L(k^2)$.
The explicit expression for $ [\calP_{\mu\nu}^L(k)]\Sigma^{ghost}_{\mu\nu}(\Omega_n,\bk)]$ in the above equations   are given by Eqs.~(\ref{PmunuLGhost})-(\ref{integrals}).
For the remaining terms, further
summations over the Lorentz indices $\mu,\nu,\alpha,\beta,\rho$ and $\sigma$ in the gluon loop, see Fig.~\ref{dse} and
 in Eqs.~(\ref{DSETrans}) and (\ref{DSELong}),
and subsequent integration over the  angle $\xi$  of terms enclosed in curly brackets  result in extremely long and cumbersome expressions. The corresponding   summation
over $\mu,\nu,\alpha,\beta,\rho$ and $\sigma$ procedure and angular integrations over $\xi$, are rather   trivial.
Nonetheless, because of the cumbersomeness of the  final expressions, in the present paper  we do not present  the resulting system  of tDSE   explicitly. We note only that the corresponding calculations
can be essentially relieved by using a symbol manipulations package, e.g. Maple 18 or Wolfram Mathematica 13. Also note, that   the
resulting expressions  for the gluon loop contain
the same, as in the previous case, integrals $\Cia_n(\alpha)$ and $\CalK_n(\alpha,b)$, Eqs.~(\ref{IntaAndIntb}), now with $n$ from $ 0$   to $n=5$.
\section{Results and discussions}\label{results}
 The   resulting   system of equations for the ghost and gluon dressing functions, Eqs.~(\ref{DSE2}), (\ref{DSETrans})-(\ref{DSELong})
represents the sought  rainbow approximation (\ref{rain2})-(\ref{rain1}) of the truncated Dyson-Schwinger equations (\ref{sdeGluon})-(\ref{sdeGhost}).
In Euclidean space,  it consists of  a system of linear algebraic equations with respect to the Matsubara frequency and  one-dimensional integral equations with respect to the three-momentum $|\bq|$. The solution of this system is
found numerically by utilisation of an iteration procedure.
 To this end, the integral equation  is replaced by  a Gaussian quadrature formula  converting  the
   Eqs.~(\ref{DSE2}),  (\ref{DSETrans})-(\ref{DSELong})
 in to  a system of algebraic equations with respect to the corresponding  Gaussian nodes and Matsubara frequencies, ready for the mentioned numerical procedure.
 The effective parameters
   in (\ref{ff1})-(\ref{ff3})  generally can be different for the transversal $(T)$ and longitudinal $(L)$ parts.  In addition, all   parameters, including the renormalization constants, $\tilde Z_3$ and $Z_3$,
 may depend on the temperature $T$ and Matsubara frequency $\Omega_n$, see Ref.~\cite{Maas:2005hs}. However,
 as it has been observed in Refs.~\cite{S-XQINprd84,Maasprd75,lattice1}, at low temperatures the interaction kernel is insensitive to the temperature impact and, as a first approximation, the interaction kernels can be chosen the same as
 at $T=0$ with $D^T=D^L$ and $\omega_i^T=\omega_i^L$. For larger temperatures such a choice of
 transversal and longitudinal effective parameters is less justified.  In the present paper we use the same values of the effective  parameters for longitudinal and transversal parts. The concrete values correspond to  the set of parameters previously found by fitting   the solution of the tDSE in vacuum, as reported in Ref.~\cite{ourGlueball}:
%
 $D_{11}^T=D_{11}^L=0.462$~GeV$^2$, $D_{12}^T=D_{12}^L=0.116$~GeV$^2$,
$\omega_{11}^T=\omega_{11}^L=1.095$~GeV, $\omega_{12}^T=\omega_{12}^L=2.15$~GeV
for the 3-gluon loop,
$D_{21}^T=D_{11}^L=7.7$~GeV$^2$, $D_{22}^T=D_{22}^L=0.25$~GeV$^2$,
$\omega_{21}^T=\omega_{21}^L=0.58$~GeV, $\omega_{22}^T=\omega_{22}^L=4.5$~GeV,
for the ghost loop and
$\omega_{31}^T=\omega_{31}^L=0.73$~GeV, $\omega_{32}^T=\omega_{32}^L=2.16$~GeV
$D_{31}^T=D_{31}^L=0.39\pi$~GeV$^2$, $D_{32}^T=D_{32}^L=0.1\pi$~GeV$^2$ for the ghost-gluon loop
(cf. the three loops in  Fig.~\ref{dse}).
\footnote{Note that in the present paper each of three loops in  Fig.~\ref{dse} have been parametraized by two Gaussian terms, while in Ref.~\cite{ourGlueball}  the  ghost loop contains only one.  The additional term does not affect the final results and was used solely to preserve uniformity in the parameterization of loops.
Correspondingly,  the present notation  differs from the one in Ref.~\cite{ourGlueball}.}

It should be stressed  that keeping the effective parameters the same as in vacuum, one can expect
reliable solution of the tDSE only at low and moderate  temperatures, $T\lesssim 200$~MeV.
For higher temperatures, the effective parameters must become $T$-  and $n$-dependent
 (cf.~Refs.~\cite{dor,S-XQINprd84} for $T$-dependence of the effective parameters and
 Ref.~\cite{Maas:2005hs} for $n$-dependence of the renormalization constants $Z_3$ and $T$-dependence of the
 coupling constant $g$) and, probably, quite different for the
  transverse and longitudinal propagators.

 We solve the mentioned system of algebraic equation numerically by an iteration procedure.
 The number of Matsubara frequencies has been fixed in the interval $(-30\,\le n\,  \le 30$), the Gaussian mesh for the spatial integration on $|\bk|^2$ consisted on 72 nodes with  the upper limit of integration  $|\bk|^2 \sim 220$~GeV$^2$/c$^2$. To obtain more dense Gaussian nodes at low $|\bk|^2$, an appropriate mapping  has been employed.
Notice that, in concrete numerical calculations one can encounter problems  evaluating
the integrands at $|\bq|\simeq |\bk|$, i.e. at $ b=\frac{|\bq|^2+ |\bk|^2}{2|\bq| |\bk|}\to 1$ when the integrals
$\CalK_n(a,b)$, Eq.~(\ref{IntaAndIntb}), diverge. A careful and scrupulous inspection of the integrands in the vicinity
 of $b=1$  shows that independently on $n$
the integrals $\CalK_n(a,b)$ contain the same kind of singularities, viz.  singularities of the type~$\sim\ln(b-1)+Ei(1,2a)$.  They were handled by employing the
Taylor expansion  of the corresponding expressions about $b=1$  with the conclusion  that
in  the final results there is a full cancelation of all the singularities. Consequently, the final expressions are
finite.
 \begin{figure}[!ht]
\begin{center}
\includegraphics[scale=0.3 ,angle=0]{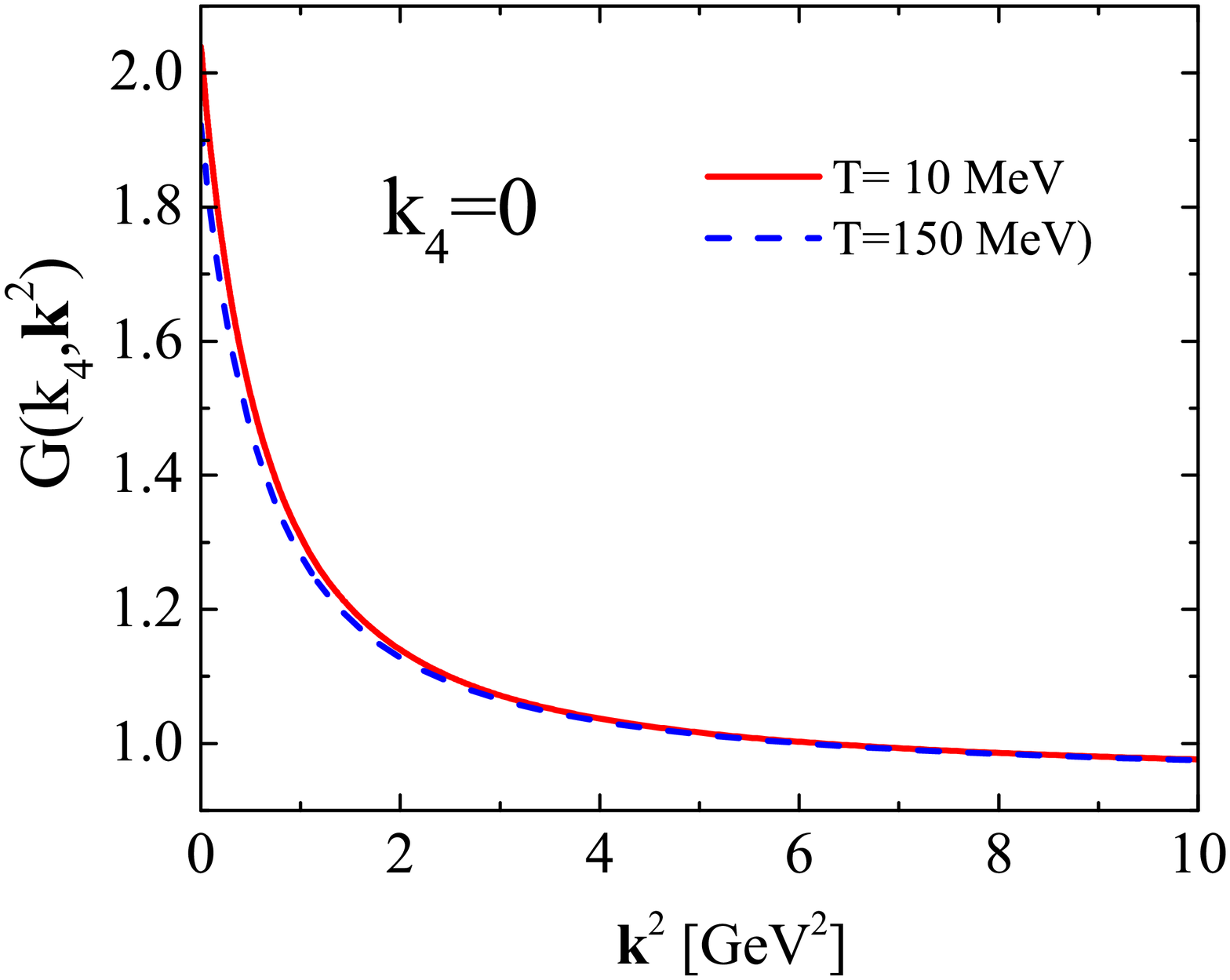}
\includegraphics[scale=0.3 ,angle=0]{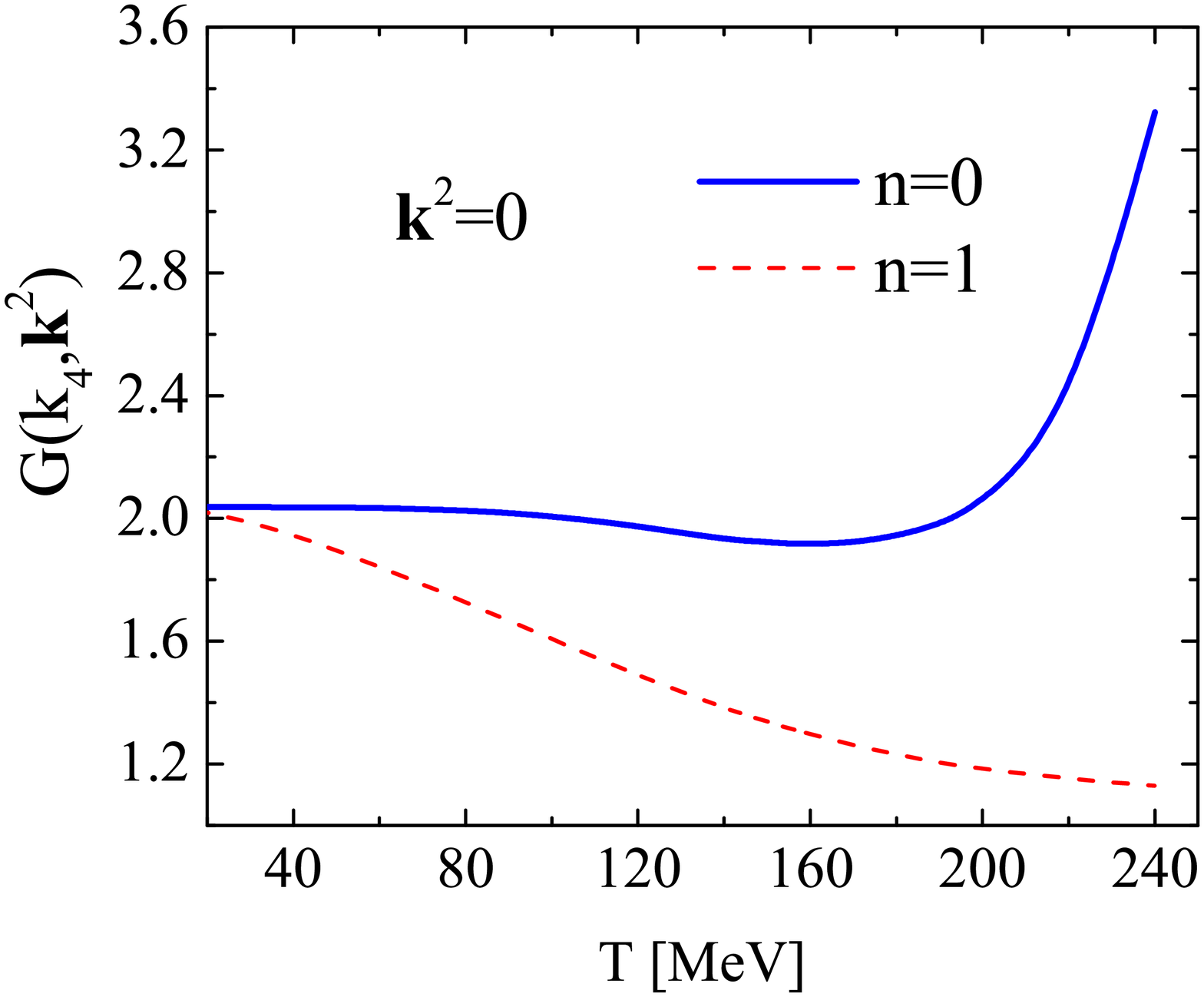}
\end{center}
 \caption{(Color online) The solution of the Dyson-Schwinger equation for the ghost dressing function $G(k_4,{\bf k}^2)$. Left panel: dependence on
 the three-momentum squared $\bk^2$
 at zero Matsubara frequency, $n=0$, for two values of the temperature $T=10$~MeV (solid curve) and $T=150$~Mev (dashed curve).
  Right panel:    temperature dependence of the  dressing function at zero three-momentum $\bk^2=0$ and
 two values of the Mastubara frequency,  $n=0$ (solid curve) and  $n=1$ (dashed curve).   }
 \label{FigGhost}
 \end{figure}
\subsection{Solution for the  the ghost dressing function $G(k_4, \bk^2)$}\label{resGhost}

 Results of calculations are presented in Fig.~\ref{FigGhost} where  we exhibit the temperature dependence of the ghost dressing function $G(k_4,\bk^2)$. Left panel illustrates the dressing function at zero Matsubara frequency, $n=0$, as a function  of three-momentum $\bk^2$ at two values of the temperature, $T=10$~MeV and $T=150$~MeV. As expected, the temperature dependence is rather weak for all values of $\bk^2$, except for the vicinity of very low momenta  $\bk^2\sim 0$ and is quite similar to the behaviour of $G(k^2)$ at $T=0$, cf. Ref~\cite{ourGlueball}.
 In the right panel we present the temperature dependence of the dressing function $G(k_4,\bk^2)$ at $\bk^2=0$ and two values of the Matsubara frequencies, $n=0$ and $n=1$. Since the fourth component of the  ghost momentum   strongly depends on the Matsubara frequency,
 $k_4^2 = [ 2n \pi T]^2$, the dressing function $G(k_4, 0)$ is also quite sensitive to~$T$ and $n$.
 This qualitatively agrees with the lattice QCD calculations reported in Ref.~\cite{Aouane:2011fv,Maasprd75,lattice1}
 where it has been found that at low enough three-momenta $|\bk|$ and $T/T_c < 1.4$ the ghost dressing function $G(k_4,\bk)$  changes rather weakly with the temperature at $k_4=0$ and   decreases as $k_4$ increases.
As seen from Fig.~\ref{FigGhost}, the dressing function is basically temperature independent, i.e.
 $\displaystyle\frac{d}{d T}G(0,0)  \sim 0$, up to $T=T_0 \simeq 150$~MeV. At $T >T_0 $ the function changes its concavity
  (the second derivative w.r.t. temperature changes the sign) and sharply increases. In some sense, $T_0$ can be considered as the critical point for the temperature dependence of the ghost dressing $G(0, 0)$. The ghost dressing for the non-zero Matsubara modes smoothly decreases with $T$ and does not exhibit any irregularities.
\subsection{Solution for the transversal and longitudinal  gluon propagators}\label{resGluont}
\begin{figure}[!ht]
\begin{center}
\includegraphics[scale=0.30 ,angle=0]{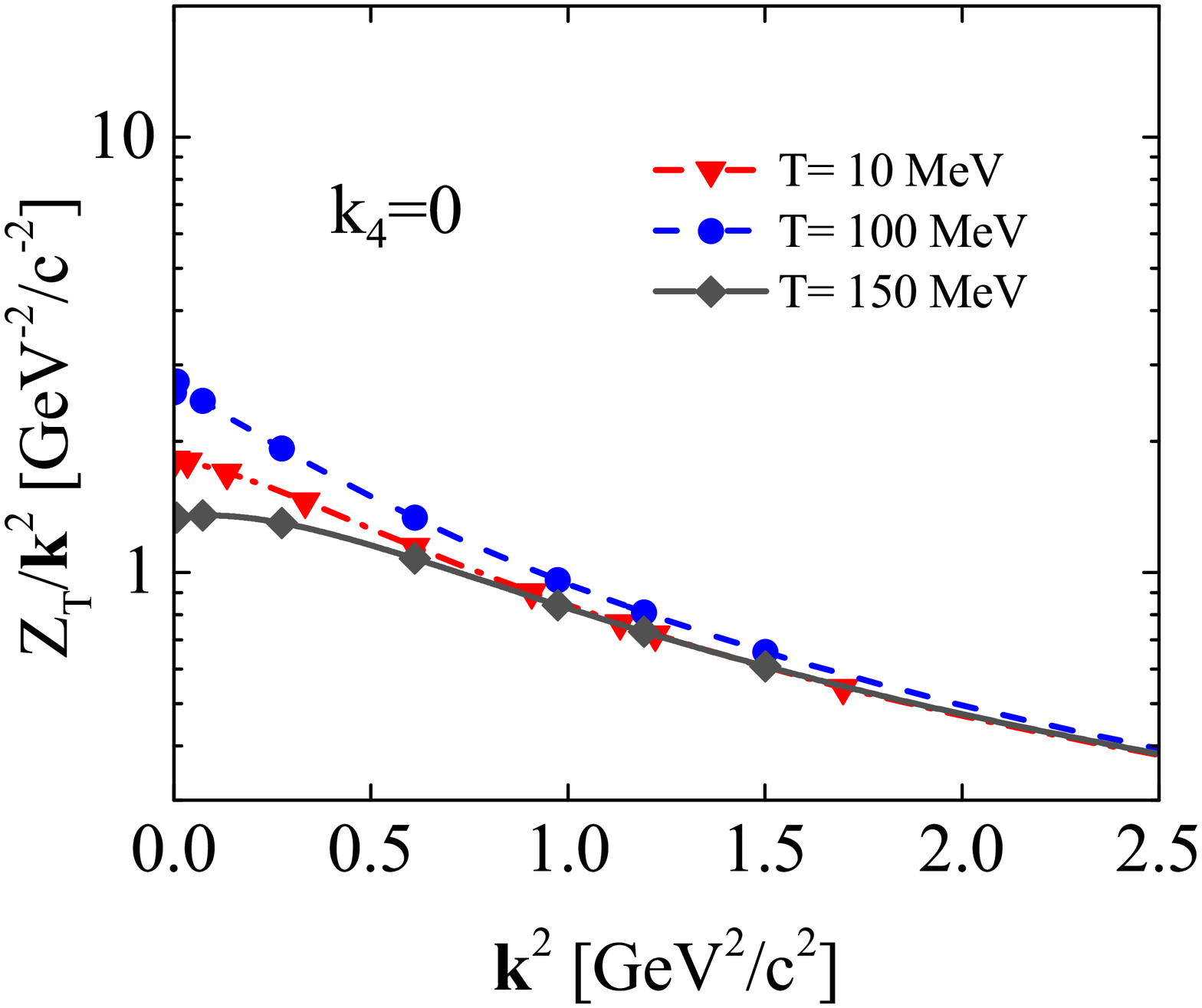}
\includegraphics[scale=0.3 ,angle=0]{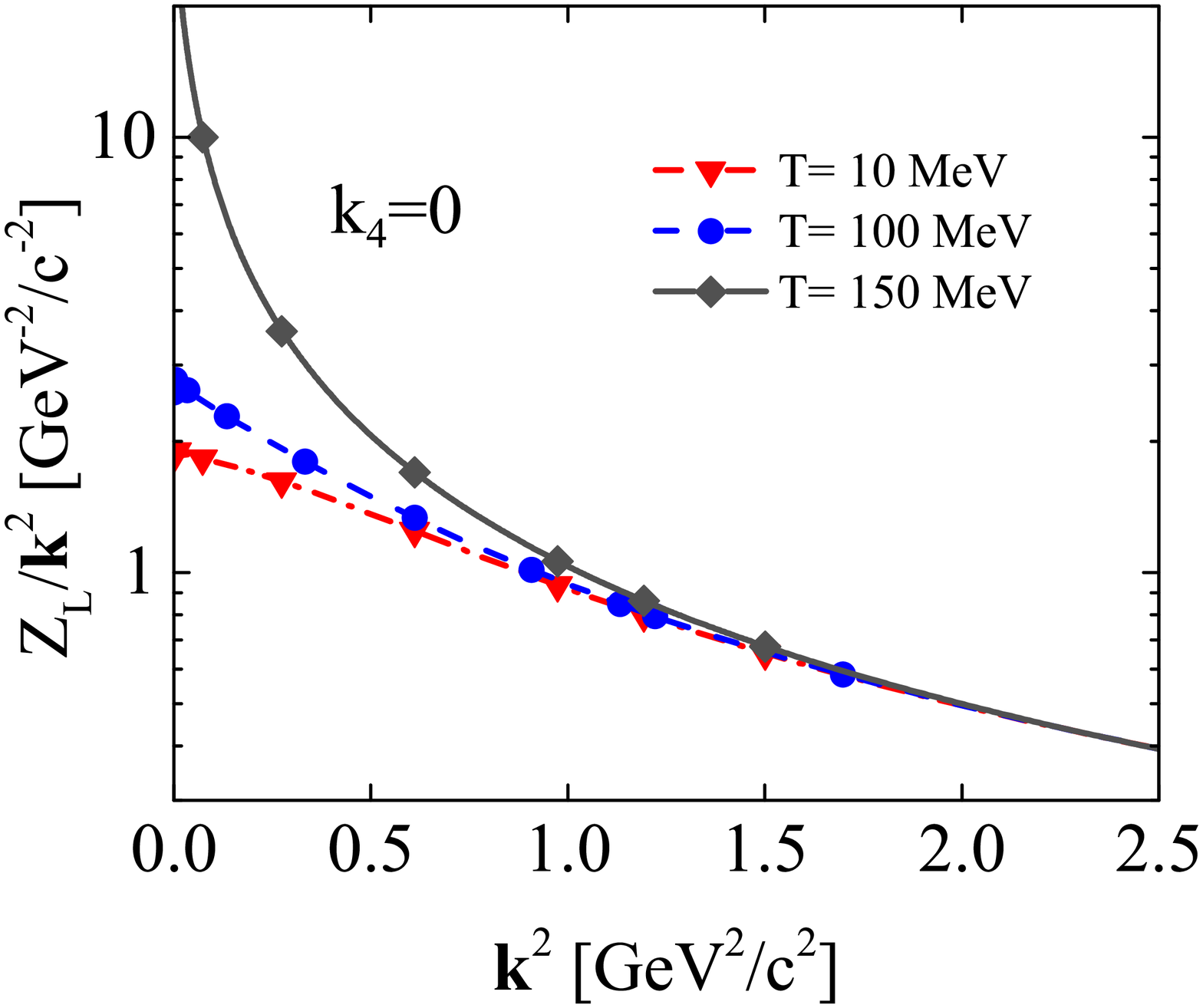}
\end{center}
 \caption{ (Color online)  Dependence  on the three-momentum squared
  $\bk^2$ of the solution of the Dyson-Schwinger equation for the  zero Matsubara
  mode of the  transversal $Z_{T}(k_4,{\bf k}^2)/k^2$ and longitudinal $Z_{L}(k_4,{\bf k}^2)/k^2$ gluon propagators, left  and right panels, respectively.
  Solid,     dashed and dot-dashed curves correspond to $T=150$~MeV, $T=100$~MeV and $T=10$~MeV respectively.
    }
\label{FigGluon}
 \end{figure}
 In Figs.~\ref{FigGluon} and \ref{FigGluon1} a similar  analysis is presented, but now for the gluon propagator.
  In Fig.~\ref{FigGluon}, the dependence of the  propagators for the  zero Matsubara mode is displayed as a function
 of the spatial momentum squared $\bk^2$ at several values of the temperature,
  $T=10$~MeV, $T=100$~MeV and $T=150$~MeV. As in the previous case,  the dependence of the
   transversal, left panel, and longitudinal, right panel, propagators  at moderate and large values of $\bk^2$ is rather weak.  The $T$-dependence  of the propagators is more pronounced in the region of low momenta $\bk^2$, where
   transversal and longitudinal propagators manifest quite different behaviour as functions of $T$. While the transversal propagator is not so sensitive to $T$, the longitudinal one sharply increases with increase of $T$.
   This is more evidently seen   in Fig.~\ref{FigGluon1}, left panel, where temperature dependence of the Matsubara zero mode is presented at $\bk^2\sim 0$. The steep behaviour  at  $T\sim T_0$ of the longitudinal propagator is manifested more distinctly. Moreover, we observed that for larger temperatures $T>T_0$, viz. $ T\gtrsim  160$~MeV,  the iteration procedure does not longer converge and the solution of the tDSE for the gluon propagators disappears.
   This is a direct indication that our approach with the temperature independent effective parameters cannot be
   extended to large temperatures.
 %
\begin{figure}[!ht]
\includegraphics[scale=0.3 ,angle=0]{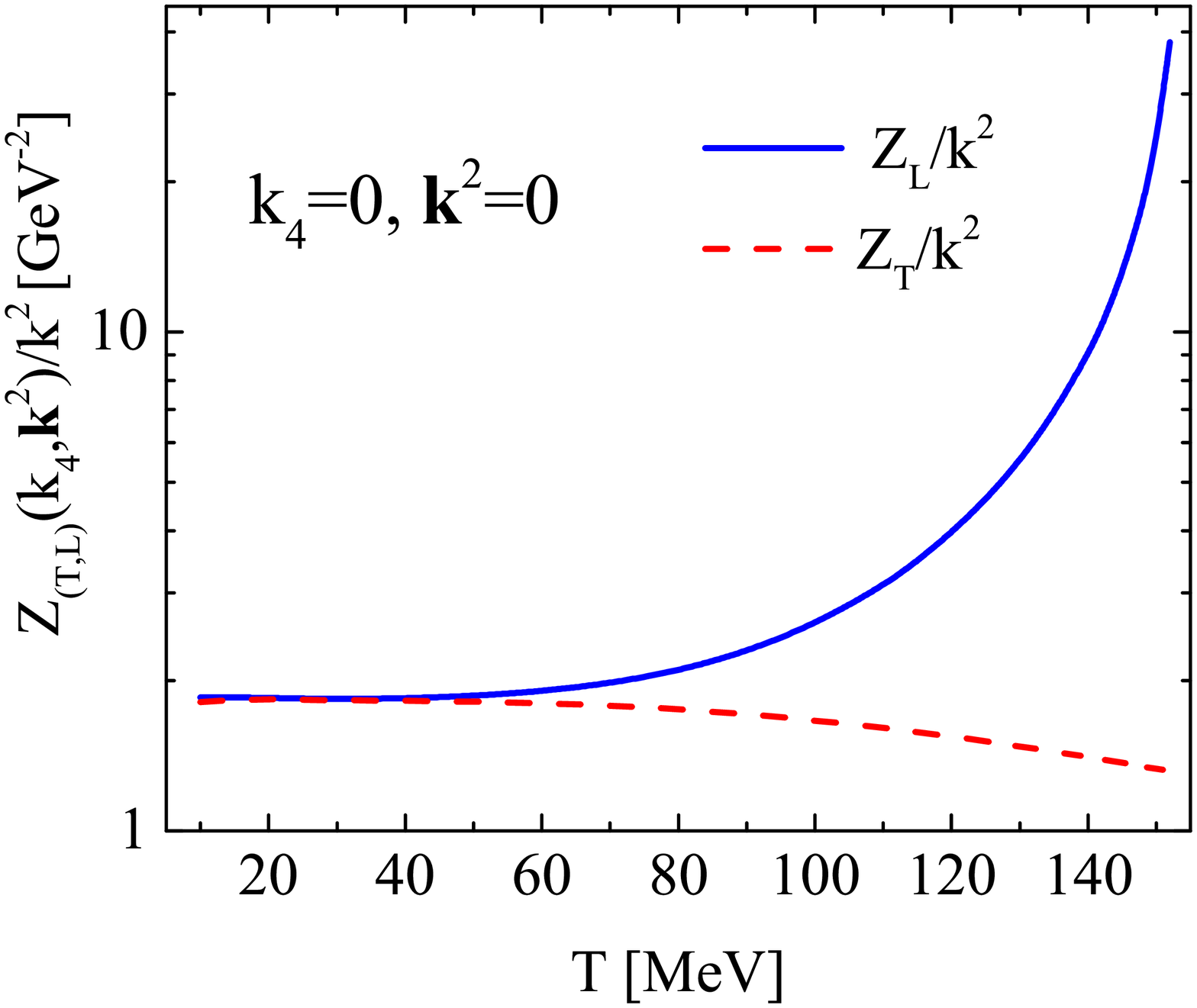}
\includegraphics[scale=0.30 ,angle=0]{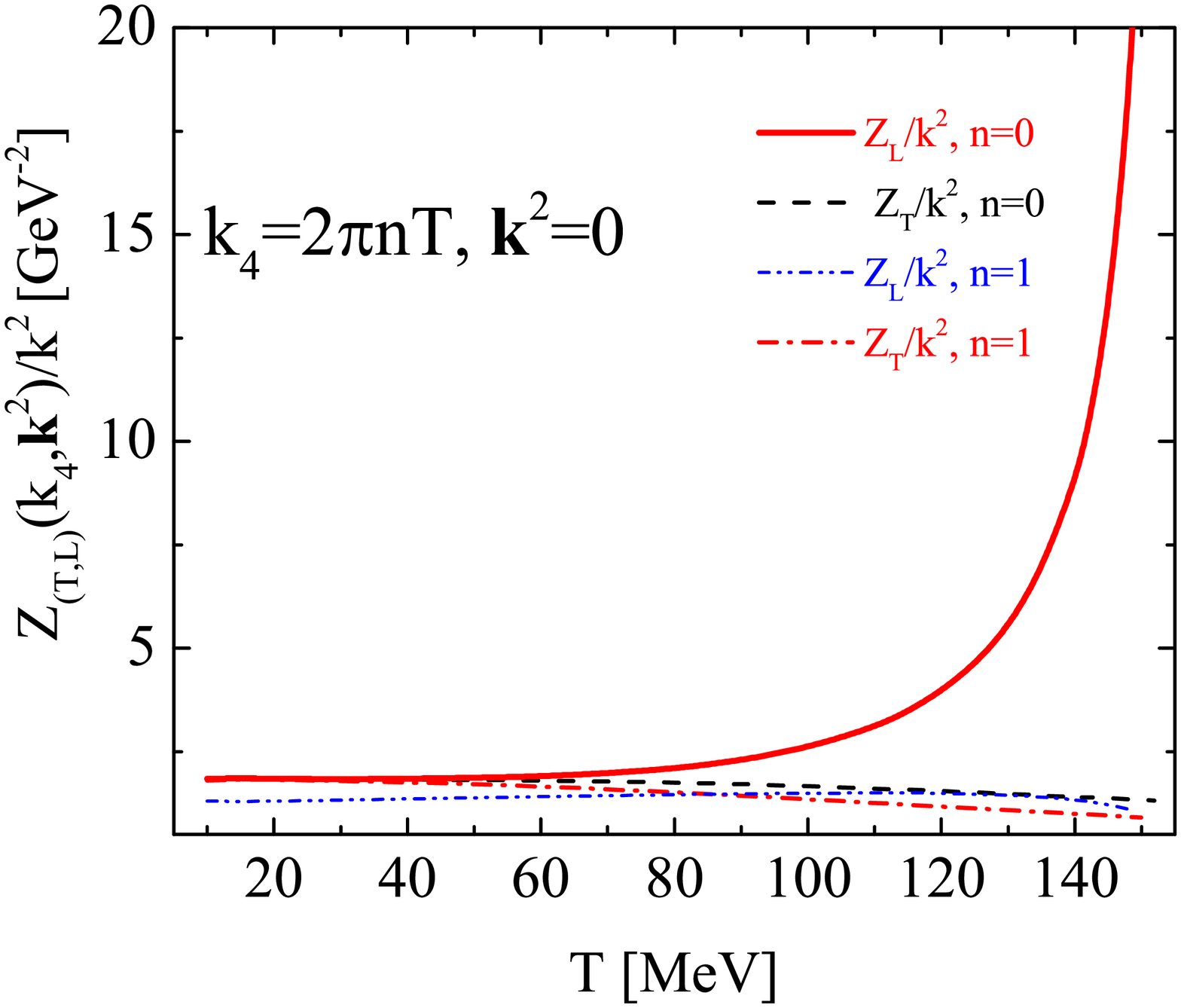}
 \caption{ (Color online) The  solution of the Dyson-Schwinger equation for the gluon propagators
  $Z_{L,T}(k_4,{\bf k}^2)/k^2$. Left panel: temperature dependence of zero Matsubara modes
  of the propagators at $\bk^2=0$. Solid curve corresponds to the longitudinal
 propagator, the dashed curve is for the transversal one. Right panel: illustration of the temperature dependence of the transversal and longitudinal propagators  for zero and non-zero Matsubara modes. Solid and dashed lines are those displayed in the left panel, i.e. for $n=0$. The dot-dot-dashed and dot-dashed lines correspond to longitudinal and transversal propagators at $n=1$, respectively.
 Note the $log$ scale on the left panel and the linear scale on the right panel.  }
\label{FigGluon1}
 \end{figure}
   The right panel in  Fig.~\ref{FigGluon1} reflects
  the $T$-dependence of the gluon propagator at $\bk^2=0$ and two values of the Matsubara frequency, $n=0$ and $n=1$.
  The different behaviour of the longitudinal propagator at moderate and high temperatures is due to
 vanishing  forth component
    $k_4^2=0$ for $n=0$ and relatively large  values of $k_4$,   $k_4^2=4\pi^2 T^2$, for $n=1$.
     Recall that  in the tDS equations with the interaction kernels (\ref{ff1})-(\ref{ff3}),
    the fourth components $k_4$ enter as $\exp(-k_4^2/\omega^2)$ which essentially affect
    the values of the corresponding propagators at finite $n$ and large $T$.  Since the transverse propagator is less sensitive to the temperature, see left panel in Fig.~\ref{FigGluon1},
   the dependence on the Matsubara modes is not so strong.
   This is in a fairly good qualitative agreement
    with the results obtained for the interval
   temperature interval ($0 \le T   \le 150$~MeV) within the FRG framework
    (see Refs.~\cite{Dupuis:2020fhh,DressingPawlowsky,PawlowskyFRG,MaasPhysReport}
    and references therein quoted) and within the  lattice QCD
    simulations~\cite{Aouane:2011fv,Ilgenfritz:2017kkp,lattice1}, where a noticeable sensitivity of the longitudinal component of the gluon propagator with respect to the deconfining phase transition has been observed.
   This effect easily see from the behaviour of the electric screening  mass
       \begin{equation}
       m_L=\left. \sqrt{\frac{k^2}{Z_L(k4,|\bk|)}} \ \right|_{k^2\to 0}.
   \label{scrM}
       \end{equation}
       Figure~\ref{screenM} illustrates the temperature dependence of $m_L$. It can be seen that
       the screening mass  is a decreasing function of $T$ in the whole, available in the present analysis,
     temperature   range $T<T_0$ in an  agreement with
       SU(2) and SU(3) lattice calculations~\cite{lattice1}. For larger temperature the lattice calculations predict
       a drastic increase of $m_L$ at $T>T_c$, where $T_c$ is a critical temperature for phase transitions
       (e.g. for chiral phase transition $T_c=154\pm 9$~MeV~\cite{critical1}, whereas the
         SU(3) pure Yang-Mills theory predicts a first order transition at the critical temperature
         $T_c \sim 270$~MeV~\cite{critical}).
  \begin{figure}[!ht]
\includegraphics[scale=0.3 ,angle=0]{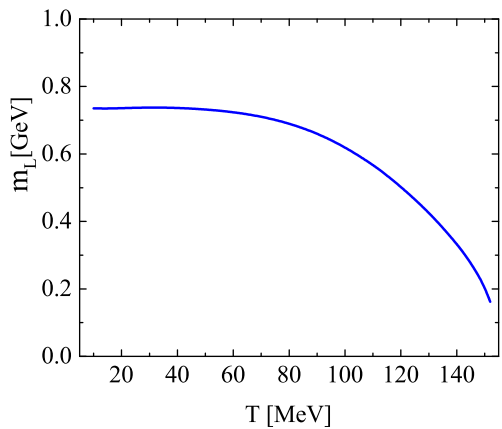}
 \caption{ (Color online)  The electric screening mass, Eq.~(\ref{scrM}), as function
 of the temperature $T$.  }
\label{screenM}
 \end{figure}

    We reiterate  that  our simplified approach with fixed parameters
    from  vacuum   cannot be reliably applied at larger temperatures,
    $T \gtrsim 160$~MeV, and a comparison with FRG and/or lattice
    QCD approaches at higher temperatures is  therefore hindered. An analysis of the possible
    temperature dependence of the effective parameters of the approach will be presented elsewhere.

  \section{Summary}\label{summary}

In summary, we solve numerically
the system of truncated gluon-ghost Dyson-Schwinger equations within the rainbow approximation at finite temperatures within the framework of the Matsubara imaginary-time formalism. The effective parameters of the interaction
kernels have been considered as temperature independent and taken from  the  previous fit of the gluon and ghost propagators to the SU(2) lattice QCD results in vacuum. We argue that for the zero-Matsubara frequency, $n=0$, the dependence of the dressing functions  on the three-momentum squared
$\bk^2$  is not sensitive to the temperature $T$ in almost the whole range of $\bk^2$. Contrarily, a strong dependence
on $n$ and $T$  has been found at the origin, $\bk^2=0$. We found that, with effective parameters fixed from vacuum calculations the iteration procedure of solving the tDSE   does not converge at large enough temperatures, $T > T_0$, where $T_0 \sim  150-160$ MeV. In the vicinity of $T_0$, the longitudinal gluon propagator sharply increases, whereas the transversal one does not exhibit irregularities. It means that, if one searches for some gluon  phase transitions,  then
the most appropriate quantity to study is the temperature dependence of the longitudinal propagators.
The presented  investigations can be considered as a first step in  further, more reliable, analysis of the $T$-dependence of gluon and ghosts propagators. The obtained solution of the tDSE can be directly implemented
 in to the  Bethe--Salpeter equations  to perform  studies of
behaviour  glueballs as two-gluon bound states in hot and dense matter.
\section{Acknowledgments}
A bulk of numerical calculations have been performed   on the basis of the HybriLIT heterogeneous
computing platform~\cite{govorun} (supercomputer "Govorun", LIT, JINR).

\end{document}